\documentclass[twocolumn,preprintnumbers,,aip,apl]{revtex4}

\usepackage{graphicx,amsmath,amssymb}

\begin{document}

\title{Negative terahertz conductivity  in disordered graphene bilayers with population inversion}

\author{D. Svintsov,$^1$ T. Otsuji,$^2$, V. Mitin,$^{2,3}$, M. S. Shur,$^4$ and V. Ryzhii \footnote{Electronic mail: v-ryzhii(at)riec.tohoku.ac.jp}
$^{2}$}
\affiliation{$^1$ Laboratory of Nanooptics and Plasmonics, Moscow Institute of Physics and Technology, Dolgoprudny 141700, Russia\\
$^2$ Research Institute of Electrical Communication, Tohoku University,  Sendai 980-8577, Japan\\
$^3$ Department of Electrical Engineering, University at Buffalo, SUNY, Buffalo, NY 1460-1920, USA\\
$^4$ Departments of Electrical, Electronics, and Systems Engineering and Physics, Applied Physics, and Astronomy, Rensselaer Polytechnic Institute, Troy, NY 12180, USA\\
}

%

%
\begin{abstract}
The gapless energy band spectra make the structures based on graphene and graphene bilayers with the population inversion
created by optical or injection pumping to be 
 promising media  for the interband terahertz (THz) lasing.
However, a strong intraband absorption at THz frequencies still poses a challenge for efficient THz lasing. 
In this paper, we show that in the pumped  graphene bilayer structures, 
the indirect interband radiative transitions accompanied by scattering of carriers caused by  disorder 
can provide a substantial negative  contribution to  the THz conductivity  (together with the direct 
interband transitions).
 In the graphene bilayer structures  on high-$\kappa$ substrates with point charged defects, these transitions  almost fully compensate the losses due to the  intraband (Drude) absorption. 
We also demonstrate that the indirect interband contribution to the THz conductivity in a graphene bilayer with the extended defects (such as the charged impurity clusters, surface corrugation, and nanoholes)
can surpass by several times the fundamental limit  associated with the direct interband 
transitions  and the Drude conductivity. These predictions   can affect the strategy of the graphene-based THz laser implementation. 
\end{abstract}

\maketitle
\newpage

\newpage

The absence of a band gap in the atomically thin carbon structures,such as  graphene and graphene bilayers, enables their applications in different terahertz (THz) and infrared devices~\cite{1,2,3,4}. One of the most challenging and promising problems is the creation of the graphene-based THz lasers~\cite{5,6,7}. These lasers are expected to operate at room temperature, particularly, in the 6-10 THz range, where the operation of III-V quantum cascade lasers is 
hindered by the optical phonons~\cite{8}. 
Recent pump-probe spectroscopy experiments confirm the possibility of the coherent radiation amplification in the optically pumped graphene~\cite{9,10,11,12,13,14,15,16}, enabled by a relatively long-living  interband population inversion~\cite{17}. As opposed to optically pumped graphene lasers, graphene-based injection lasers are expected to operate in the continuous mode, with the interband population inversion maintained by the electron and hole injection from the $n$- and $p$-type contacts~\cite{18}. 
A single pumped graphene sheet as the  gain medium provides   the maximum radiation amplification coefficient 
corresponding to
 the quantity $4\pi \sigma_Q/c = \pi \alpha = 2.3\%$, where $\sigma_Q = e^2/4\hbar$
is the universal optical conductivity of a single graphene layer, $e$ is the electron charge, $\hbar$ is the Planck constant and  $\alpha \simeq 1/137$ is the fine-structure constant~\cite{17}. 
The THz gain  in  the graphene bilayers~\cite{19} or the non-Bernal stacked multiple-graphene layers~\cite{5}  can be enhanced  approximately proportional to the number of the layers. However, more crucial is 
 the problem of competition between the interband radiation amplfication and intraband (Drude) radiation absorption~\cite{17,20}. The latter scales with frequency $\omega$ approximately as $1/(1 +\omega^2\tau^2)$, where $\tau$ is the momentum relaxation time. Hence, the onset of terahertz gain is typically believed to occur only in clean samples, where $\omega \tau \gg 1$~\cite{21}.

\begin{figure}[t]
\centering
\includegraphics[width=7.0cm]{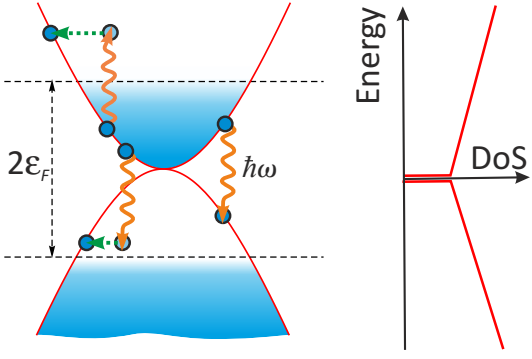}
\caption{Energy band diagram of a graphene bilayer with population inversion (left) and schematic energy dependence of its density of states (right). Direct interband transitions involve only vertical photon emission (wavy arrow), while indirect emission and absorption  processes are accompanied
by  carrier scattering (straight horizontal arrows).}
\label{F1}
\end{figure}

In this letter, we show that the  intraband radiation absorption in disordered graphene bilayers does not actually pose a problem for the THz lasing. On the contrary, the presence of a certain type of defects reinforces the negative contribution to the  THz conductivity and, hence,
improves the gain properties of the pumped graphene bilayers. The origin of such an effect is
associated with
the  indirect interband radiative transitions (see Fig.~1), in which
the  electrons from the conduction band emit  the photons being scattered  on the defects,  thus contributing to the radiation gain. The inclusion of such processes can be important in the indirect-gap materials and gapless semiconductors. However, in single-layer 
graphene  with charged impurities, the indirect interband processes contribute to the net  THz conductivity  $\sigma$ only moderately due to a low  density of states (DoS) near the band edges (except, possibly, for the case of the carrier interactions with  artificial 
fairly large-size scatterers \cite{22}) . 
The situation is remarkably different is the intrinsic graphene bilayers, where the electron-hole dispersion is gapless and almost parabolic~\cite{23}. With constant DoS near the band edges, the indirect interband radiative processes significantly contribute to the net radiation gain under the population inversion conditions. 
The THz gain  is proportional to $- {\rm Re}~\sigma$.
We  analyze the spectral dependencies of the real part of the net dynamic THz conductivity 
${\rm {Re}}~\sigma$  in the pumped graphene bilayers for the  scattering by the point charged impurities~\cite{24,25} and by
the impurity clusters~\cite{26,27}. Our analysis  shows that in the first case, the gain due to the indirect interband transition almost fully compensates the losses due to the Drude absorption. 
In the case of the cluster scattering,  the gain he indirect interband transitions  exceeds  
the intraband losses. 
Depending on the cluster size and their density, the ratio $|\rm {Re}~\sigma|/\sigma_Q$ can markedly exceed unity resulting elevated net THz gain.

The real part of the net in-plane dynamic THz conductivity, ${\rm Re}~\sigma$, comprises the contributions of the direct (vertical) interband electron transitions,  ${\rm Re}~\sigma_d$, and the contributions, ${\rm Re}~\sigma^{intra}_{id}$ and ${\rm Re}~\sigma^{inter}_{id}$, of 
two types of the indirect electron transitions (inside both the bands and 
 between the two bands), respectively:
 
\begin{equation}\label{eq1}
{\rm Re}~\sigma = {\rm Re}~\sigma_d + {\rm Re}~\sigma^{inter}_{ind} + {\rm Re}~\sigma^{intra}_{ind}.
\end{equation}

The interband conductivity of graphene bilayers due to the direct transitions is given by~\cite{19,23}

\begin{multline}
\label{Eq2}
 {\rm Re}~\sigma_d = \sigma_Q \frac{\hbar\omega + 2\gamma_1}{\hbar\omega +\gamma_1}
\left[f_v(-\hbar\omega/2) - f_c(\hbar\omega/2)\right]\\
= \sigma_Q \frac{\hbar\omega + 2\gamma_1}{\hbar\omega +\gamma_1}
\tanh\biggl(\frac{\hbar\omega - 2\varepsilon_F}{4T} \biggr),
\end{multline}
where $f_v(\varepsilon)$ and $f_c(\varepsilon)$ are the carrier distribution functions in the valence and conduction bands,  and $\gamma_1 \simeq 0.4$~eV is the hopping integral between carbon atoms in adjacent graphene planes~\cite{28}. Here and in the following, 
for the intrisic graphene bilayers
we assume $f_v$ and $f_c$ are the Fermi functions characterized by the quasi-Fermi energies $\mu_v = -\varepsilon_F$ and $\mu_c = \varepsilon_F$, respectively. This is justified by strong carrier-carrier scattering leading to a fast  thermalization of excited carriers~\cite{9,13}. In the THz frequency range of interest, $\hbar \omega \ll \gamma_1$, and the conductivivty of bilayer is simply twice as large as that for a single layer.  
Under the pumping conditions, $\mu_c = - \mu_v = \varepsilon_F > 0$, and, according to Eq.~(2), the THz conductivity  is negative for photon energies $\hbar \omega$ below the double quasi-Fermi energy $\varepsilon$ of pumped carriers~\cite{17}.

To evaluate the real part of the THz conductivity due to the indirect transitions, we calculate the second-order transition amplitudes for the photon emission (absorption) accompanied by the electron scattering and apply the Fermi's golden rule. As a result, the general expressions for the indirect intra- and interband THz conductivities read as follows:

\begin{multline}
\label{Eq3}
{\rm Re}~\sigma_{ind}^{intra} =  \frac{8 \pi \sigma_Q}{\hbar\omega^3}
\sum_{{\bf k}, {\bf k}^{\prime},\lambda}
[f_{\lambda}(\varepsilon_{{\bf k}}) - f_{\lambda}(\varepsilon_{{\bf k}^{\prime}})]\\ 
\times\delta(\hbar\omega + \varepsilon_{{\bf k}} - \varepsilon_{{\bf k}^{\prime}})
|V_S({\bf k} - {\bf k}^{\prime})|^2 u^{\lambda\lambda}_{{\bf k}{\bf k}^{\prime}}({\bf v}_{{\bf k}^{\prime}} -
{\bf v}_{{\bf k}})^2,
\end{multline}

\begin{multline}
\label{Eq4}
{\rm Re}~\sigma_{ind}^{inter} =  \frac{8 \pi \sigma_Q}{\hbar\omega^3}
\sum_{{\bf k}, {\bf k}^{\prime}}
[f_v(\varepsilon_{{\bf k}}) - f_c(\varepsilon_{{\bf k}^{\prime}})] \\
\times\delta(\hbar\omega - \varepsilon_{{\bf k}} - \varepsilon_{{\bf k}^{\prime}})
|V_S({\bf k} - {\bf k}^{\prime})|^2 u^{cv}_{{\bf k}{\bf k}^{\prime}}({\bf v}_{{\bf k}^{\prime}} +
{\bf v}_{{\bf k}})^2.
\end{multline}
Here $\lambda =c, v$ is the index corresponding to the conduction and valence bands, ${\bf v}_{{\bf k}} =2\hbar v_0^2 {\bf k}/\gamma_1$ is the electron velocity in the graphene bilayer,  $v_0 \simeq 10^8$cm/s is the velocity characterizing the energy spectra of graphene and graphene bilayers.$V_S({\bf q})$ is the ${\bf q}$-th Fourier component of the scattering potential, and $u^{\lambda\lambda}_{{\bf k}{\bf k}^{\prime}} = (1 + \lambda\lambda^{\prime}\cos \theta_{{\bf k}{\bf k}^{\prime}})$ is the overlap between the envelope wave functions in graphene bilayers. Equation~(3) reproduces the well-known result for the dynamic conductivity in the high-frequency limit $\omega \gg \tau^{-1}$, where $\tau$ is the momentum relaxation time, which can be derived from the Boltzmann equation. For a correct qualitative description of the low-frequency conductivity, we replace the frequency $\omega$ in the denominators of Eqs.~(2) and (3) with $(\omega^2 + \tau^{-2})^{1/2}$.

In the most practical situations,the Coulomb scattering by the random substrate-induced charged impurities is the main factor determining the conductivity~\cite{29}. Considering impurities as random uncorrelated point scatterers with the average density $n_i$, we write the scattering matrix element as~\cite{30}
  
\begin{equation}\label{Eq5}
|V_S({\bf q})|^2 = n_i\biggl[\frac{2\pi e^2}{\kappa(q + q_s)}\biggr]^2,
\end{equation}
where $\kappa$ is the background dielectric constant, and $q_s$ is the Thomas-Fermi screening wave vector~\cite{30}.
In the case of the interband population inversion, the latter is found to be $q_s = 4 \alpha_c \gamma_1(1 - \ln 2)/\hbar\,v_0$, where $\alpha_c = e^2/\hbar\kappa\,v_0$ is the coupling constant.
  Evaluating the integral, we find the scattering-assisted interband conductivity

\begin{figure}[t]
\centering
\includegraphics[width=6.0cm]{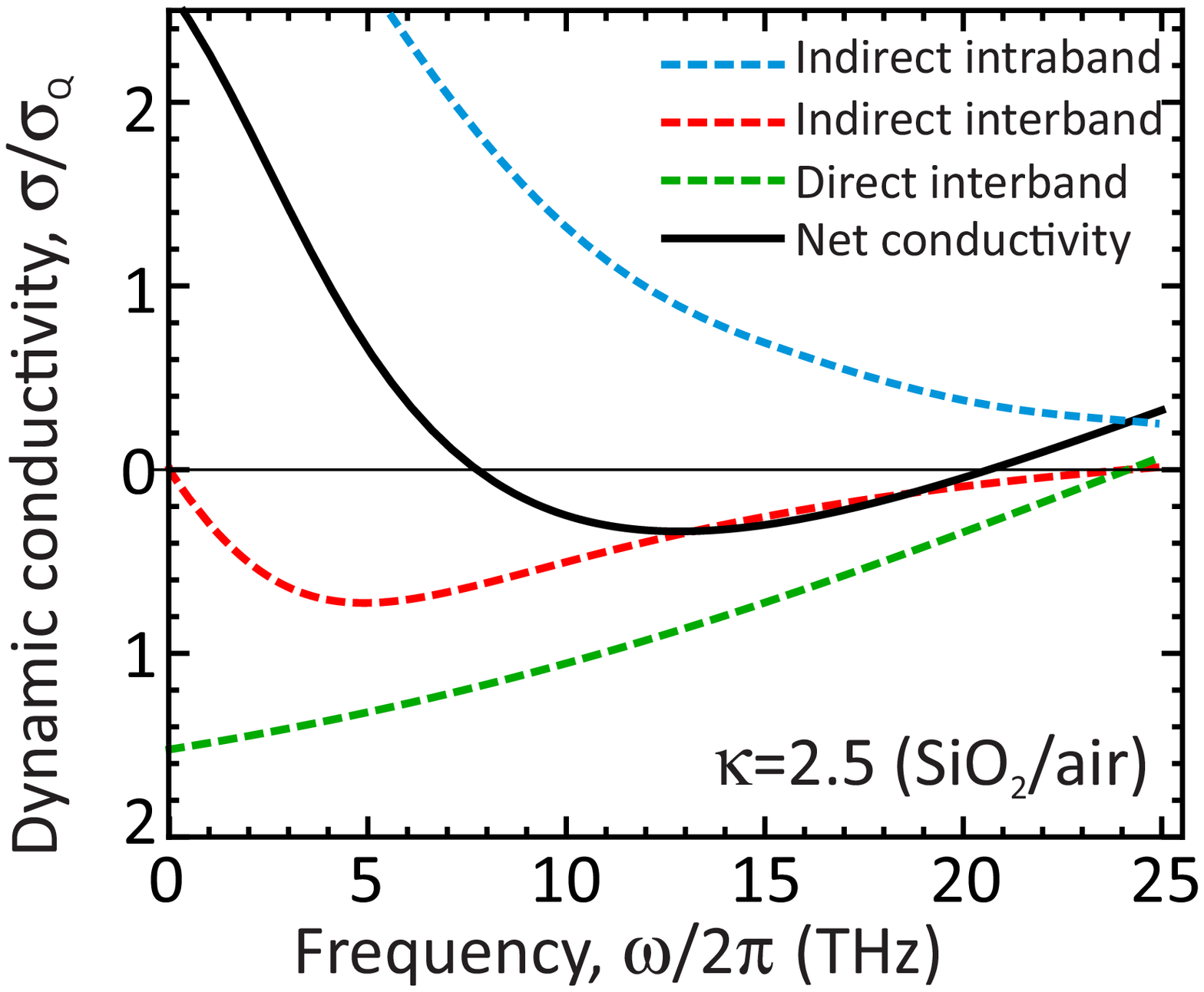}
\includegraphics[width=5.8cm]{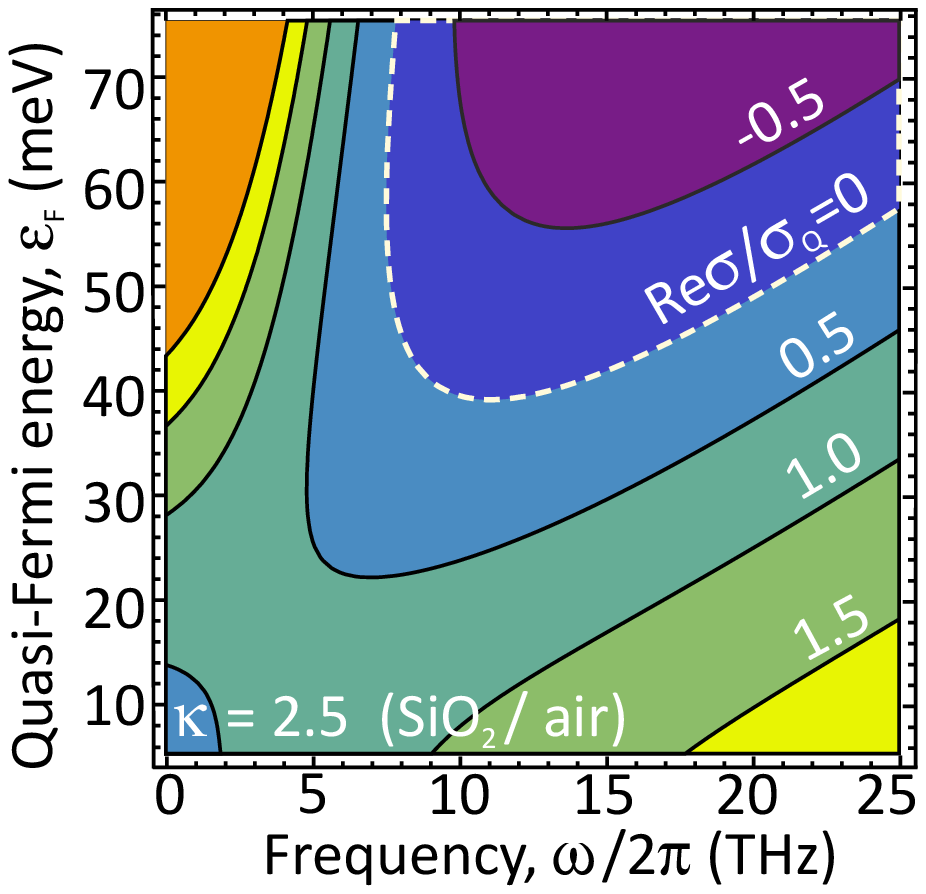}
\caption{Contributions of different radiative transitions  to   dynamic THz conductivity of a graphene bilayer on SiO$_2$ ($\kappa = 2.5$) and its net value (solid line) versus frequency
for  $n_i=10^{12}$ cm$^{-2}$ and  $\varepsilon_F = 50$ meV  (upper panel),
and $\varepsilon_F - \omega$ color map (bottom panel). Numbers above the lines indicate the values of ${\rm Re}~\sigma/\sigma_Q$. The region above  dashed line corresponds to the negative  conductivity.}
\label{F2}
\end{figure}

\begin{figure}[th]
\centering
\includegraphics[width=6.0cm]{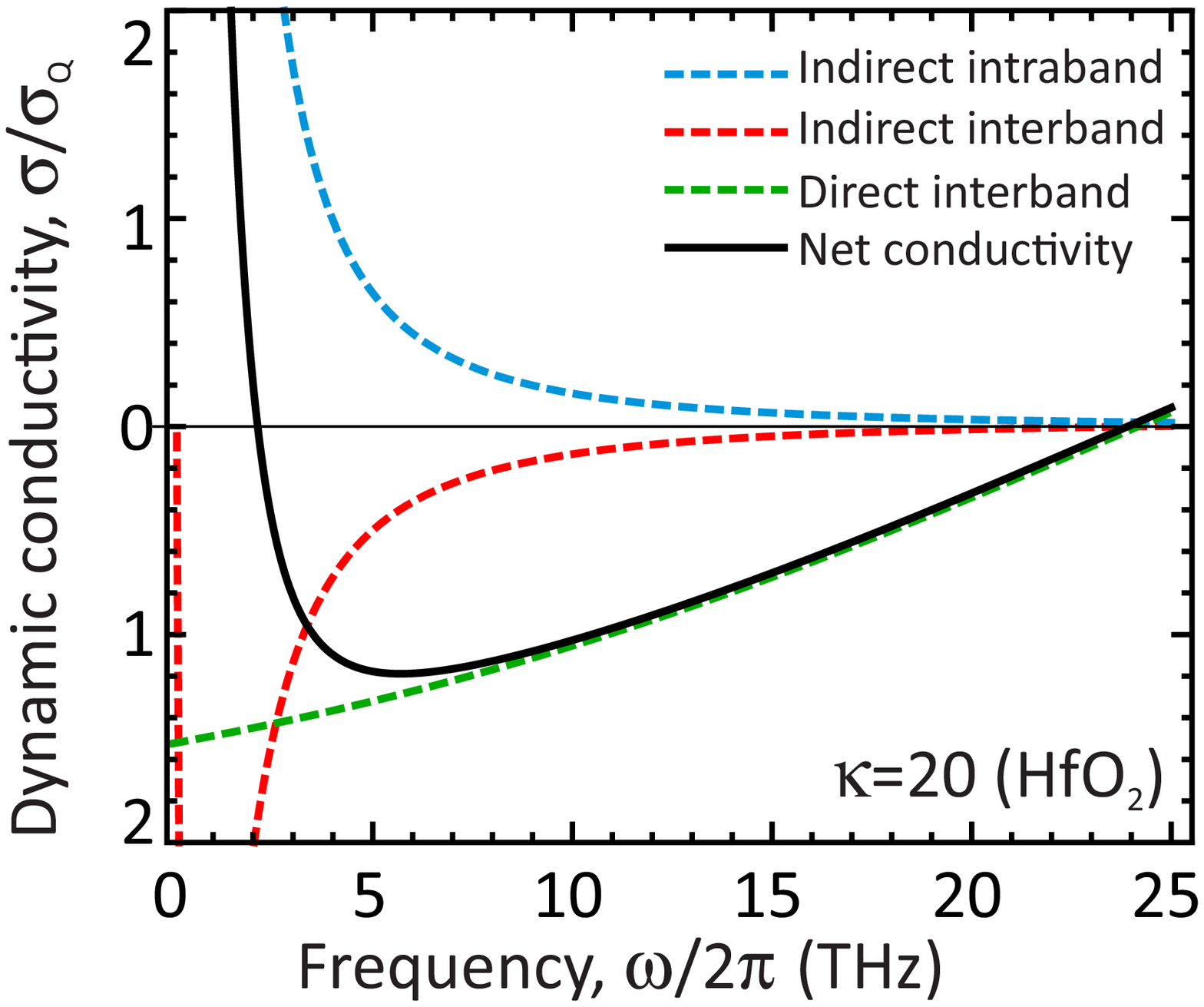}
\includegraphics[width=5.8cm]{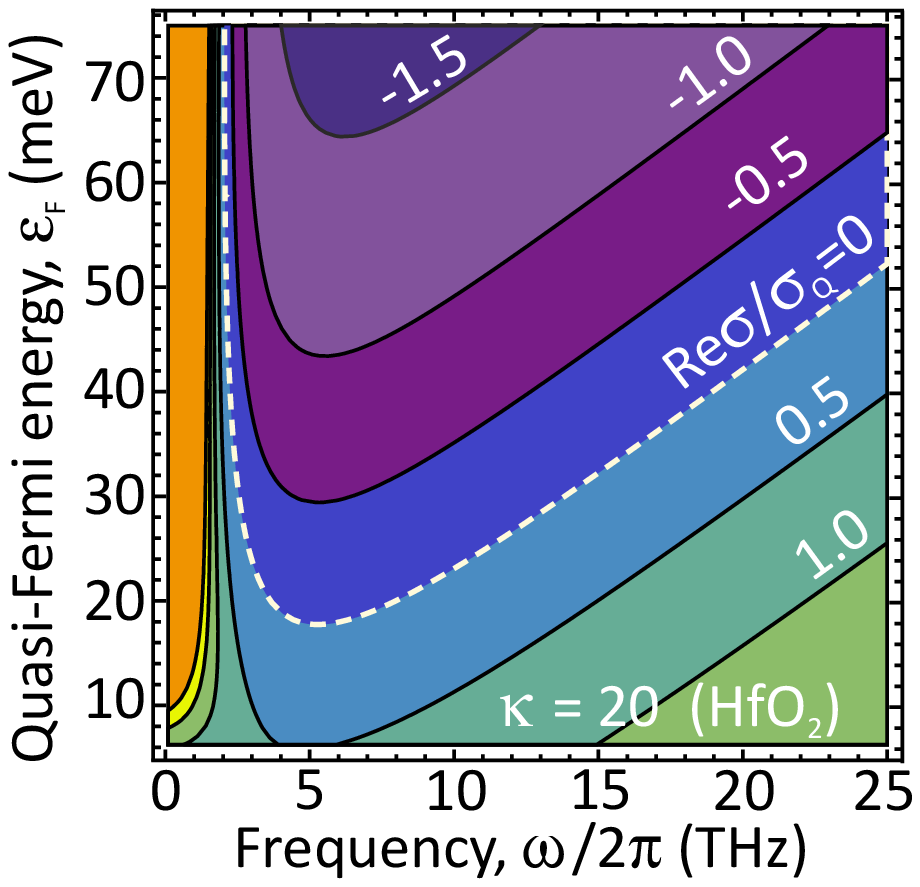}
\caption{The same as in Fig.~2 but for graphene bilayer clad by Hf0$_2$ layers ($\kappa = 20$).}
\label{F3}
\end{figure}

\begin{multline}
\label{Eq4}
{\rm Re}~\sigma_{ind}^{inter} \simeq 8 \pi \alpha_c^2 \sigma_Q \frac{v_0^2 n_i}{(\omega^2 + \tau^{-2})}\\ 
\times
 \tanh\biggl(\frac{\hbar\omega - 2\varepsilon_F}{4T} \biggr)\Phi\biggl( \frac{\hbar v_0 q_s} {\sqrt{\hbar \omega \gamma_1}}\biggr),
\end{multline}
where $\Phi(x) = \int_0^{2} dy (1 - |1-y|) (1-y/2)/(y^{1/2} + x)^2$ with the following  asymptotes: $\Phi(x) \approx 2 \ln 2 -1/2$, $x \ll 1$, $\Phi(x) \approx 1/(2x^2)$, $x \gg 1$. 
At intermediate frequencies, $1/\tau \ll \omega \ll \hbar (v_0 q_s)^2/\gamma_1$, the conductivity due to the indirect interband transitions scales as $\omega^{-1}$. This is substantially  different from the case of a single graphene layer, where it tends to a constant. A pronounced increase in the indirect interband contribution to the dynamic conductivity in a bilayer at low frequencies is attributed to the constant density of states in the vicinity of the band edges. 

Figure~2 (upper panel) shows the spectral dependencies of the contributions of different radiative transitions
to the dynamic THz conductivity as well the net value of the latter
  for a moderate-quality exfoliated graphene bilayer ~\cite{25} on the
SiO$_2$ substrate
with the effective dielectric constant $\kappa = 2.5$ 
(graphene bilayer sandwiched between SiO$_2$ and air) impurity density $n_i = 10^{12}$~cm$^{-2}$, at temperature $T = 300$~K, and for the quasi-Fermi energy $\varepsilon_F = 50 $~meV. The bottom panel in Fig.~2 demonstrates the two-dimensional maps of the dynamic THz conductivity versus the 
 frequency and the quasi-Fermi energy for the same structure.
 As seen, the contribution of the indirect interband transitions in such a sample
 is weaker than those of both the indirect intraband  and direct interband
 transitions, although the former is essential  for compensating, to some extend,  the Drude absorption.
 This enables an increase in  $|{\rm Re}\sigma|$ and some widening of the frequency range where
 ${\rm Re}~\sigma < 0$.
 As follows from Fig.~2 (bottom panel),
 the dynamic THz conductivity is  negative at $\varepsilon_F \gtrsim 40$~meV and 
 $\omega/2\pi \gtrsim 10$~THz.
 
 The relative contribution of the indirect interband transitions(compared to the "normal" intraband conductivity) increases with an increasing  dielectric constant. 
 Figure~3 shows the same characteristics as in Fig.~2,  for a graphene bilayer being
 clad by the HfO$_2$ layers, i.e., immersed  in a media with a fairly high dielectric constant ($\kappa = 20$). As seen from Fig.~3, the transition to the graphene bilayer structures with higher dielectric constant leads to a much stronger contribution of the indirect interband transitions
 and, hence, to a much larger value of $|{\rm Re}~\sigma|$. The comparison of  Figs. 2 and 3
 shows that the ratio $|{\rm Re}~\sigma|/\sigma_Q$  for  the minimum  value of of $\sigma$
 at $\kappa = 20$ is five-six time larger than for $\kappa = 2.5$. Moreover, in the former case
  ${\rm Re}~\sigma$ becomes negative starting from $\varepsilon_F = 20$~meV
 and  $\omega/2\pi \gtrsim 5$~THz. At higher values of $\varepsilon_F$ (i.e., at a stronger pumping),
 ${\rm Re}~\sigma$ can be negative from the frequencies of about a few THz to a dozen THz.

 The  reinforcement of the negative dynamic THz conductivity effect 
 with increasing background dielectric constant
 demonstrated above is interpreted as follows.
An increase in $\kappa$ results in the reduced   Thomas-Fermi wave vector $q_s$.
This, in turn, leads to switching from the strongly screened to almost bare Coulomb scattering. From the energy conservation laws it follows that the indirect interband electron transitions are favored by a low momentum transfer $q$, while for the indirect transitions within one band $q$ should be large, namely, $q > \omega / v_0$ as the electron velocity in GBL does not exceed $v_0$. Hence, for high values of $\kappa$, the scattering potential behaves approximately as $V_S({\bf q}) \propto 1 / q$, which supports the interband transitions with a low momentum transfer.

\begin{figure}[t]
\centering
\includegraphics[width=7.5cm]{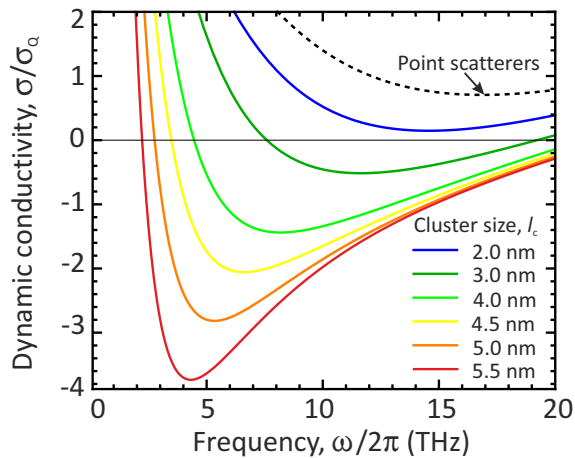}
\caption{Net  THz conductivity vesus frequency calculated for graphene bilayers on SiO$_2$
with impurity clusters of different size $l_c$: cluster  density and charge are 
$n_c = 10^{11}$~cm$^{-2}$ and $Z = 7$, dielectric constant $\kappa=2.5$, and quasi-Fermi energy  $\varepsilon_F = 50$~meV. Dashed line corresponds to  point charged defects with the same charge and density.}
\end{figure}

As shown above, the Coulomb scattering by the screened point defects cannot lead to the dominance of the indirect interband transitions over the indirect intraband ones. However, such a dominance can be realized in the case of the carrier scattering on sufficiently extended scatterers.
We consider  the charged cluster of size $l_c$ and charge $Z e$ as a continuous two-dimensional distribution of charge density $\rho({\bf{r}}) = Z e \exp(-r^2/l_c^2)/\pi l_c^2$ (since the exact density distribution is not very important, for simplicity, we assume it to be Gaussian~\cite{27}). Solving the Poisson equation for this charge density and averaging over the random positions of the clusters, we readily find the scattering matrix elements  [instead of Eq.~(5)]

\begin{equation}\label{Eq7}
|V_S({\bf q})|^2 = n_c \biggl[\frac{2\pi Z e^2 }{\kappa(q + q_s)}\biggr]^2 \exp\biggl( - \frac{q^2 l_c^2}{2}\biggr).
\end{equation}
Here  $n_c$ is the density of the clusters. 
In the limit of strong screening, $q \ll q_s$, our model of scattering coincides with the widely accepted model of the "Gaussian correlated disorder"~\cite{31,32,33,34} with the root-mean-square scattering potential $\sqrt{{\overline {V^2}}} \approx 2\pi \sqrt{n_c/\pi l_c^2} Z e^2/\kappa q_s$.

Figure~4 shows the spectral characteristics of the dynamic THz conductivity in the pumped graphene bilayers ($\varepsilon_F =50$~meV) with clusters of charged impurities with different size $l_c$. If the scattering by the charged clusters is a dominating mechanism, the absolute value of the net  dynamic THz conductivity $|{\rm Re}~\sigma|$can markedly exceed
the fundamental "direct interband limit" of $2\sigma_Q$. At reasonable values of $l_c\gtrsim 6 $ nm and quasi-Fermi energy $\varepsilon_F = 50$ meV, one obtains
$|{\rm Re}~\sigma| > 4\sigma_Q$. 

In conclusion, we have demonstrated that the 
graphene bilayers with a long-range disorder (impurity clusters with  reasonable values of the cluster density and size), can exhibit a strong negative  THz conductivity 
with the span two times or more   exceeding the fundamental limit  $2\sigma_Q$.
 This effect is associated with the indirect interband  transitions with the photon emission being accompanied  by the disorder scattering. For the indirect interband photon emission to dominate over the "normal" Drude absorption, the long-wave length  Fourier components of the scattering potential should prevail. Such kind of scattering potentials can be formed also by  extended surface corrugations, quantum dots on the graphene bilayer surface~\cite{35},   and nanohole arrays~\cite{36,37}.

The work was supported by 
the Japan Society for Promotion of Science (Grant-in-Aid for Specially Promoted Research $\#$ 23000008) and by 
the Russian Scientific Foundation (Project $\#$14-29-00277) and the grant of the Russian Foundation of Basic Research $\#$ 14-07-31315. 
The works at UP and RPI were  supported  by the US Air Force
award $\#$ FA9550-10-1-391  and by the US Army Research Laboratory
Cooperative Research Agreement, respectively.
 

\end{document}